\pdfoutput=1

\documentclass[11pt]{article}

\usepackage[final]{acl}

\usepackage{times}
\usepackage{latexsym}

\usepackage[T1]{fontenc}

\usepackage[utf8]{inputenc}

\usepackage{microtype}

\usepackage{inconsolata}

\usepackage{graphicx}

\usepackage{booktabs}

\title{A Multi-Agent Dual Dialogue System\\to Support Mental Health Care Providers}

\author{
    Onno P. Kampman\textsuperscript{1},
    Ye Sheng Phang\textsuperscript{1},
    Stanley Han\textsuperscript{1},
    Michael Xing\textsuperscript{1},
\\
    \textbf{Xinyi Hong\textsuperscript{1}},
    \textbf{Hazirah Hoosainsah\textsuperscript{1}},
    \textbf{Caleb Tan\textsuperscript{1}},
    \textbf{Genta Indra Winata\textsuperscript{2}\thanks{The work was done outside Capital One.}},
\\
    \textbf{Skyler Wang\textsuperscript{3}},
    \textbf{Creighton Heaukulani\textsuperscript{1}},
    \textbf{Janice Huiqin Weng\textsuperscript{1}},
    \textbf{Robert JT Morris\textsuperscript{1}}
\\
\\
    \textsuperscript{1}MOH Office for Healthcare Transformation, Singapore,
    \textsuperscript{2}Capital One,
    \textsuperscript{3}McGill University
\\
\\
    Correspondence: \texttt{onno.kampman@moht.com.sg}
\\
}

\begin{document}

\maketitle

\begin{abstract}
    We introduce a general-purpose, human-in-the-loop dual dialogue system to support mental health care professionals.
    The system, co-designed with care providers, is conceptualized to assist them in interacting with care seekers rather than functioning as a fully automated dialogue system solution.
    The AI assistant within the system reduces the cognitive load of mental health care providers by proposing responses, analyzing conversations to extract pertinent themes, summarizing dialogues, and recommending localized relevant content and internet-based cognitive behavioral therapy exercises.
    These functionalities are achieved through a multi-agent system design, where each specialized, supportive agent is characterized by a large language model.
    In evaluating the multi-agent system, we focused specifically on the proposal of responses to emotionally distressed care seekers.
    We found that the proposed responses matched a reasonable human quality in demonstrating empathy, showing its appropriateness for augmenting the work of mental health care providers.
\end{abstract}

\section{Introduction}

Mental healthcare systems globally are experiencing increased pressure due to rising demands and shortages of trained professionals.
This situation has resulted in overall inadequate care and support.
Waiting lists for talk therapy, one of the main interventions for mental health conditions, are long, and therapy sessions can be prohibitively expensive for many.
Furthermore, mental health care professionals frequently experience burnout, exhibiting symptoms such as emotional exhaustion and depersonalization~\citep{oconnor2018burnout}.

Digital technology can play a crucial role in addressing these needs and alleviating the pressure on the mental healthcare system.
This paper proposes a \emph{dual dialogue system} to support therapists in conversations with their clients.
To clarify, we refer to all mental health professionals as therapists throughout this paper.
However, support systems may also include other professionals, such as counselors, psychologists, or general practitioners, as well as paraprofessionals, such as peer supporters or trained volunteers.
The proposed system is designed for general purposes with sufficient flexibility to be adapted to various domains and contexts.

\begin{figure}[t]
  \centering
  \includegraphics[width=\linewidth]{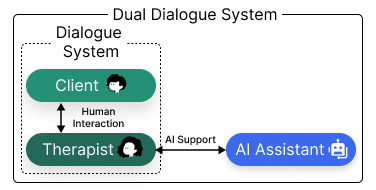}
  \caption{Schematic of dual dialogue system design.}
  \label{fig:dual-dialogue-system}
\end{figure}

Dual dialogue systems (Figure~\ref{fig:dual-dialogue-system}) can be distinguished from standard \emph{dialogue} systems because they do not directly interact with users (or \emph{clients} in a therapeutic context) but rather facilitate conversation between two humans.
They can be differentiated from \emph{trilogue} or \emph{triad} systems~\citep{Lai2023PsyLLMSU, lee2023ai} because they are not conversations between three agents; there is no communication between the client and the AI assistant. 
A system that augments human-to-human interaction instead of facilitating human-to-AI interaction in a therapeutic setting may be more suitable for real-world deployment, as it acknowledges and respects the importance of the interpersonal relationship between a care provider and a care seeker. 

In our system, the AI assistant analyzes conversations, proposes responses, recommends psychoeducational resources and self-help exercises, and evaluates and summarizes conversations for a therapist.
Importantly, the system is not intended to detect or diagnose mental illness~\citep{zhang2022natural} but to facilitate conversation by supporting therapists.
Some potential beneficial areas may include suggestions on what is troubling the user, an instant summary of lengthy or previous sessions, or the immediate generation of resources tailored to the client's needs.

Below, we first discuss the background and context of the proposed system.
The initial focus is on its application within the Singaporean mental health ecosystem.
Being specific is valuable, as it allows us to discuss a real-world application and collect grounded evidence.
However, we stress that the system may be generalized to different nations and contexts.
We then detail the requirements of the system, its design principles, and the functionalities included.
Subsequently, we describe system evaluation strategies and present a response-quality evaluation study that includes both machine and human raters.
Finally, we conclude with the system's limitations and ethical considerations.

\section{Background}

\subsection{AI and Chatbots in Mental Health}

Technology-enabled assisting approaches, such as chatbots and dialogue systems, have been proposed as ways to support and augment therapeutic labor to scale and improve mental health support~\citep{althoff2016large, Bendig2019TheNG, CaceresNajarro2023WMGPTT2, Cameron2017TowardsAC, hua2024large, jin2023psyeval, Lai2023PsyLLMSU, Lee2020IHY, li2023systematic, maddela2023training, Malgaroli2023NaturalLP, Vaidyam2019ChatbotsAC, vanHeerden2023GlobalMH}.
Such approaches may be particularly useful in cultures in which seeking professional help is still stigmatized or otherwise inaccessible.
More specifically, we consider agents in such systems as \emph{role-playing} mental health professionals~\citep{Shanahan2023RolePW}.

Therapeutic chatbots are not a new phenomenon, with a long heritage stretching back to the first chatbot ever developed, ELIZA, which mimicked conversations with a Rogerian psychotherapist~\citep{weizenbaum1966eliza}.
However, recent advances in large language models (LLMs) have led to increased interest in and hopes for developing systems with real-world applicability.

Previous studies have explored such systems~\citep{Brocki2023DeepLM, CaceresNajarro2023WMGPTT2, winata2017nora}.
Some have been brought to market as commercial products, such as \textit{Woebot}~\citep{Fitzpatrick2017DeliveringCB} and \textit{Wysa}~\citep{Inkster2018AnEC}.
Other studies have demonstrated the efficacy of fully automated internet-based cognitive behavioral therapy (iCBT) delivery~\citep{kwek2024effectiveness, lu2021randomised, Linardon2024CurrentEO}.
However, mere access to self-help resources and online CBT exercises has not been a panacea for improving care~\citep{Torous2024GenerativeAI}.
Having a human in the loop remains important for effective care and safety requirements, as most users prefer a more safeguarded experience.
Moreover, standalone and self-help apps have limited utility if they are not well-integrated into healthcare systems.

Recent studies evaluating the efficacy of LLMs in offering mental health intervention have shown promising results.
When evaluated against human responses, responses by LLMs such as GPT-4 were found to demonstrate more empathy~\citep{Luo2024empathy}, active listening, and helpfulness when responding to relationship or general health-related questions~\citep{Vowels2024helpfulness, Young-etal-2024}.
Another study indicated that GPT-4, when prompted to act as a therapist, exhibits competency, empathy, and therapeutic capacity when delivering single-session therapy to individuals seeking assistance with relationship challenges~\citep{Vowels2024LLMtherapist}.
All of this evidence suggests the potential for deploying LLMs to assist human therapists in their work.

\begin{figure*}[t]
  \centering
  \includegraphics[width=\linewidth]{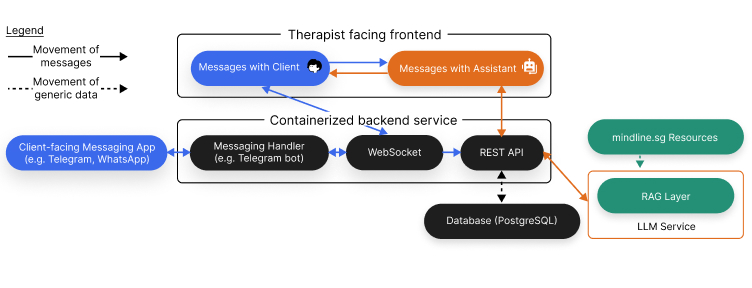}
  \caption{
    Schematic of system architecture of dual dialogue system.
  }
  \label{fig:dual-dialogue-system-architecture}
\end{figure*}

\subsection{Mental Health in Singapore}

In Singapore, the Covid-2019 pandemic has brought mental health to the fore, becoming a key priority for the government~\citep{Yeo2024PrevalenceAC}.
This has led to the establishment of national mental health support platforms, such as \texttt{mindline.sg}\footnote{\url{https://mindline.sg/}}~\citep{weng2024mental, yoon2024user} and MindSG\footnote{\url{https://www.healthhub.sg/programmes/mindsg/}}.

From this former initiative, a public Reddit-inspired forum aimed at younger Singaporeans, a demographic found to be particularly vulnerable~\citep{Subramaniam2019TrackingTM}, was born.
Users on this forum can anonymously discuss topics, provide peer support~\citep{Griffiths2012TheEO}, and ask questions to professional therapists on the platform~\citep{phang2023perceptions}.
Various other support systems and touchpoints exist in Singapore, including crisis helplines, educational content providers, community outreach programs, and private (physical and online) support providers.
However, like many other nations, Singapore experiences a shortage of mental health professionals.
In 2023, there were 298 registered psychiatrists~\citep{SingaporeMedicalCouncil2023} and 691 registered psychologists~\citep{SingaporePsychologicalSociety}.
This translates to 5.1 and 11.9 licensed psychiatrists and psychologists per 100,000 population, respectively.
The shortage of psychiatrists and psychologists results in a median waiting time of 45 days and 42 days for a subsidized appointment, respectively~\citep{MOHstats}.
Like many other nations, such an understaffed mental health workforce is often overwhelmed by workload and experiences burnout~\citep{yang_stress_2024}.

\begin{figure*}[t]
  \centering
  \includegraphics[width=\linewidth]{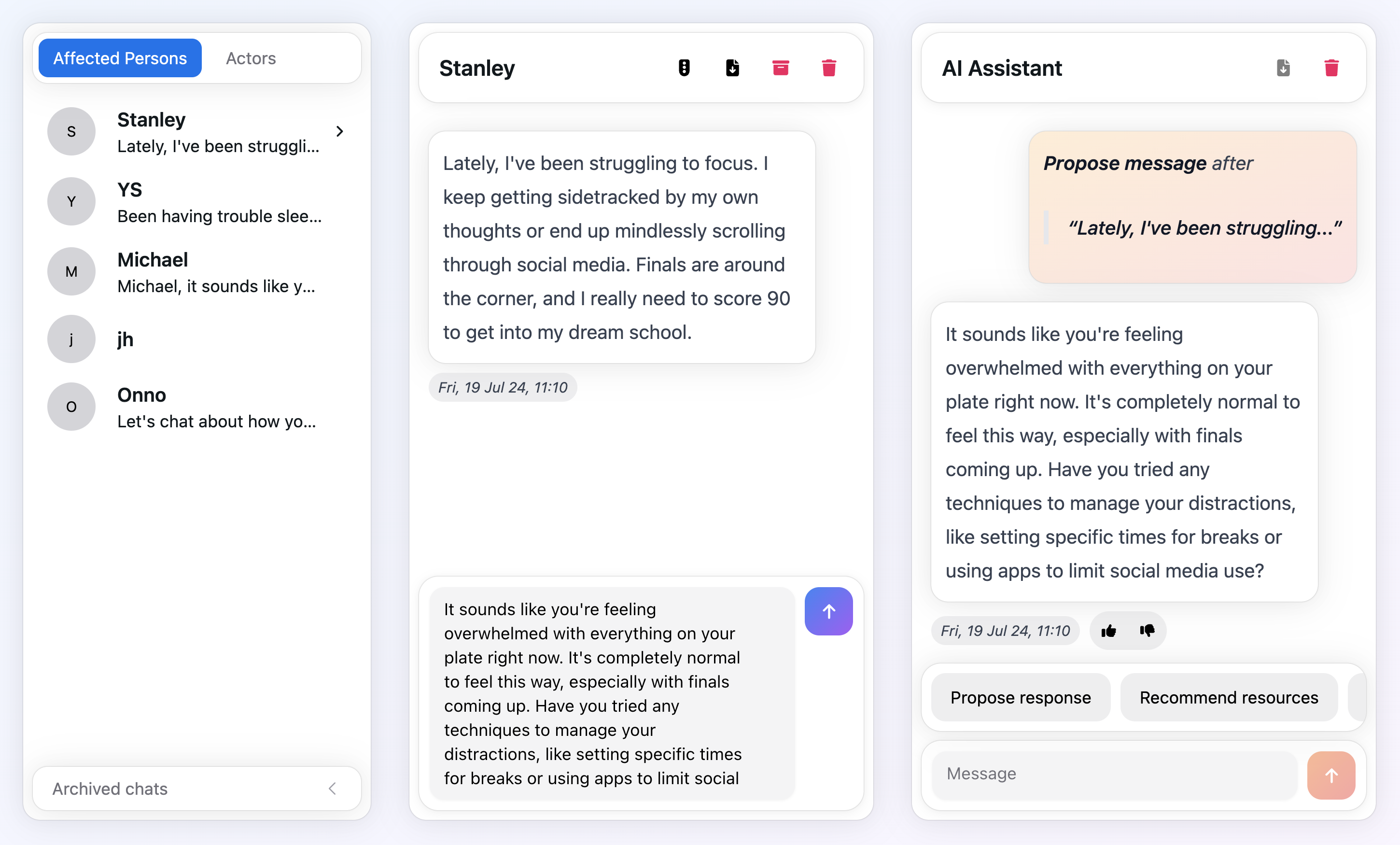}
  \caption{
    The therapist's front-end user interface showcases the two concurrent conversations and AI features, including "Propose response," "Recommend resources," and open-ended chat.
  }
  \label{fig:frontend-ui}
\end{figure*}

\section{System Description}

\subsection{System Design and Requirements}     

Our system is co-designed and co-developed with mental health stakeholders in Singapore.
Ethical considerations in the system design are important due to the sensitivity of personal data in the mental health domain.
System requirements must be such that trust is maintained for all users of the system.
We envisioned a dual dialogue system without interaction between the client and the AI assistant to minimize the risk of any LLM hallucinations directly affecting clients.
Further, we wanted the client interface to be as familiar as possible, so we drew design inspiration from popular messaging apps (Telegram or WhatsApp).
On the therapist's side, the system is designed to be general-purpose and modular; therapists may opt to use only parts of the system.
In contrast to dialogue systems that directly interact with a user, our dual dialogue system centers on therapists driving the conversation.

Natural language understanding in the AI assistant service is managed through a collection of LLM agents.
In the context of our system, an ``LLM agent'' is defined as a specialized LLM-based module designed to perform a distinct task within the broader architecture of a given mental health support system.
Each agent employs customized prompts to elicit specific types of information or responses~\citep{priyadarshana6prompt}, ensuring that its functionality is precisely aligned with its designated role.
Further, LLM agents are distinguished by their access to different types of information.
While some may tap into external databases or resources (e.g., a curated library of mental health materials), others operate solely based on the input provided by therapist and client interactions without needing external information.
This modular and specialized approach allows for a scalable, adaptable system in which each agent contributes uniquely to the holistic support framework.
The therapist is considered the ``orchestrator'' of the system agents.

Because empathy is a particularly pertinent ingredient to help achieve therapeutic outcomes~\citep{Browne-etal-commonfactors} and promote therapeutic change~\citep{Rogers1957}, it is important for our dual dialogue system to embody this quality.
Already, empathy in chatbots has been a popular topic of study and an important requirement for existing conversational systems~\citep{Fung2016TowardsEH, Ma2020ASO, Zhou2018TheDA}.
Empathy fosters a more supportive and warm environment and aids in the development of emotional connection, allowing users to feel more supported, validated, cared for, and understood both emotionally and cognitively~\citep{Decker2013TES}, ultimately facilitating client openness towards therapeutic change.
Furthermore, studies have shown that empathetic responses can drive therapeutic impacts such as developing a deeper awareness and acceptance towards personal emotions and needs and gaining better emotional regulation strategies~\citep{Irarrazaval2022Empathy}.
Commonly used LLMs already appear to be empathetic, likely through its reinforcement learning from human feedback (RLHF) fine-tuning, but we will evaluate this further in Section~\ref{sec:system-evaluation}.
Traditionally, emotionally intelligent dialogue systems would include ``affective computing'' modules, but such approaches have become less crucial in the age of generative AI~\citep{amin_will_2023}.

\subsection{System Architecture}

As shown in Figure~\ref{fig:dual-dialogue-system-architecture}, the back-end is separated from the therapist-facing front-end and the client-facing messaging platform.

The back-end infrastructure was constructed using a microservice architecture.
This architecture enables scalable interaction between the front-end and containerized backend services through well-defined API calls.
The back-end service was designed to deliver AI functionalities, such as proposing tailored responses, suggesting relevant resources, analyzing and summarizing conversations, and supporting open-ended chats with the AI assistant.
All of these rely heavily on the LLM model for its text generation and chat capabilities.
For each specific function, we crafted customized template prompts based on specific mental health needs and adjusted the model's initialization settings, such as whether to use retrieval-augmented generation (RAG) and the optimal maximum output token length.
Our interaction with the LLM service API is facilitated by LangChain~\citep{Chase_LangChain_2022} for its extensive compatibility and strong community support.
Alternatives that support RAG and could act as a drop-in replacement of LangChain include LlamaIndex~\citep{Liu_LlamaIndex_2022} and Haystack~\citep{Deepset_Haystack_2019}.

The therapist-facing front-end was built using the React JavaScript library.
The UI (see Figure~\ref{fig:frontend-ui}) was developed and co-designed alongside therapists to ensure the interface meets their needs and expectations.
It primarily consists of two chat spaces in which mental health professionals converse, allocating the space on the left for their clients and that on the right for the AI assistant.
The chat space on the left follows the design of a conventional chat interface, where the chats are listed in a side panel, and the current conversation with the client appears as a scrolling list of message bubbles.
Our system only considers text modality, but multimedia input can be included through automatic speech recognition (ASR) and multi-modal models.
The chat space on the right is where the care provider interacts with the AI assistant.
The system is designed to be flexible with a modular interface.
Therapists may only use functions of their choosing, depending on their preferred workflow and needs.
Each LLM model is tasked with one separate functionality through the multi-agent system design.
The available AI functionalities were designed based on feedback from professional therapists and include the following:

\begin{enumerate}
    \item \textbf{Propose Response}: This feature processes the entire conversation with the client to suggest an appropriate response, which can be refined before sending. Suggesting response templates in this manner may improve service quality by reducing the reply time of therapists. It relies on prompt engineering to foster engagement and promote therapeutic dialogue, focusing on eliciting clients' feelings and facilitating a deeper exploration of their experiences.
    \item \textbf{Recommend Resources}. This feature retrieves the most pertinent self-help resources for care seekers using a RAG framework. Resources are pulled from the \texttt{mindline.sg} database, including over 600 articles, videos, chatbot-guided exercises inspired by cognitive behavioral theory (CBT), and in-person opportunities for follow-up support. These curated resources cater to a broad spectrum of self-help needs. This capability significantly reduces the time therapists would otherwise spend sifting through the website to find appropriate and relevant resources. RAG also helps to avoid LLM hallucination~\citep{ji2023survey} by limiting the options of what is retrieved to a relatively small set of curated content. Specifically, we converted the descriptions of resources available on \texttt{mindline.sg} into vector formats using the GPT-based \texttt{text-embedding-ada-002} embedding model from OpenAI’s Embeddings API.
    The use of commercial APIs here was considered ethical because no personally sensitive data was included. Subsequently, we employed the Facebook AI Similarity Search (FAISS) library for efficient similarity search and clustering within this high-dimensional vector space. FAISS, designed for fast nearest-neighbor retrieval in large datasets, allows our system to rapidly identify and cluster resource descriptions based on their semantic similarity to the content of the conversation~\citep{douze2024faiss}.
    \item \textbf{Analyze Conversation}. This feature analyzes the entire conversation with the client to extract key themes, intents, and issues that emerge during the dialogue. This analysis is focused on helping the therapist find effective leverage points for psychological first aid, offering guidance on therapeutic techniques, and/or providing constructive feedback on potential instances of ineffective or inappropriate advice. By distilling the essence of conversations and suggesting follow-up actions, therapists are equipped with a potentially deeper and more nuanced understanding of clients' concerns and the therapeutic process.
    \item \textbf{Summarize Conversation}. This feature summarizes the entire conversation transcript with the client to quickly bring a therapist up to speed with the matters discussed. This is especially relevant when one therapist takes over a client from another or when a client reaches out again after an extended period of inactivity.
    \item \textbf{Empathetic Rewrite}. This feature takes a draft reply from the therapist and rewrites it in an empathetic manner. This injection of empathy follows a set of guidelines that should be adopted in the response, including engagement in active listening, usage of non-judgmental language, and validation of the client's emotions and perspectives. The conversation with the client is also included as context in the system prompt so that the rewritten response would fit the flow and style of the conversation. 
    \item \textbf{Open-ended Chat}. This functionality loads the conversation into an LLM, from which the therapist can have an open-ended chat about the conversation with the client.
\end{enumerate}

\begin{table*}[!ht]

    \centering
    \small

    \begin{tabular}{l | cccc}

        \toprule
        & \multicolumn{4}{c}{Response Generation Source} \\
        \midrule
        TES facet & Human & GPT-4o & Llama 3 70B & Llama 3 8B \\
        \midrule
        \verb|concern| &
            $4.48 \pm 1.53$ &
            $4.00 \pm 1.66$ (-1.06, 0.29) &
            $5.20 \pm 1.53$ (1.66, 0.10) &
            $5.16 \pm 0.99$ (1.87, 0.07) \\
        \verb|resonate| &
            $4.12 \pm 1.48$ &
            $4.08 \pm 1.61$ (-0.09, 0.93) &
            $4.84 \pm 1.49$ (1.72, 0.09) &
            $5.08 \pm 1.29$ (2.45, \textbf{0.02}) \\
        \verb|warmth|&
            $4.52 \pm 1.81$ &
            $3.80 \pm 1.87$ (-1.38, 0.17) &
            $5.04 \pm 1.65$ (1.06, 0.29) &
            $5.28 \pm 1.46$ (1.64, 0.11) \\
        \verb|attuned| &
            $4.08 \pm 1.63$ &
            $4.00 \pm 1.58$ (-0.18, 0.86) &
            $5.08 \pm 1.63$ (2.17, \textbf{0.04}) &
            $4.80 \pm 1.44$ (1.65, 0.10) \\
        \verb|cognitive| &
            $4.36 \pm 1.58$ &
            $3.84 \pm 1.28$ (-1.28, 0.21) &
            $4.60 \pm 1.83$ (0.50, 0.62) &
            $5.00 \pm 1.08$ (1.67, 0.10) \\
        \verb|understanding| &
            $4.44 \pm 1.83$ &
            $3.84 \pm 1.62$ (-1.23, 0.23) &
            $4.76 \pm 1.79$ (0.63, 0.54) &
            $5.00 \pm 1.55$ (1.17, 0.25) \\
        \verb|acceptance| &
            $4.80 \pm 1.66$ &
            $4.20 \pm 1.71$ (-1.26, 0.21) &
            $5.32 \pm 1.44$ (1.19, 0.24) &
            $5.44 \pm 1.36$ (1.49, 0.14) \\
        \midrule
        \verb|average| & 
            $4.40 \pm 1.65$ & 
            $3.97 \pm 1.62$ & 
            $4.98 \pm 1.62$ & 
            $5.11 \pm 1.31$ \\
        \bottomrule
        
    \end{tabular}
    
    \caption{
        Human rater perceived empathy scores (TES item ratings) for each response generation source.
        The mean $\pm$ standard deviation and $t$-test results are shown between machine-generated responses and (professional) human-written responses ($t$-value and $p$-value in brackets).
        Significant differences ($p<0.05$) are bolded.
    }
    \label{tab:eval-mean-sd-human-raters}
\end{table*}

\begin{table*}[!ht]

    \centering
    \small

    \begin{tabular}{l | cccc}

        \toprule
        & \multicolumn{4}{c}{Response Generation Source} \\
        \midrule
        TES facet & Human & GPT-4o & Llama 3 70B & Llama 3 8B \\
        \midrule
        \verb|concern| &
            $5.40 \pm 1.26$ &
            $5.48 \pm 0.51$ (0.29, 0.77) &
            $5.64 \pm 0.57$ (0.87, 0.34) &
            $5.60 \pm 0.58$ (0.72, 0.47) \\
        \verb|resonate| &
            $3.96 \pm 0.93$ &
            $4.48 \pm 0.51$ (2.44, \textbf{0.02}) &
            $4.32 \pm 0.56$ (1.65, 0.10) &
            $4.12 \pm 0.44$ (0.77, 0.47) \\
        \verb|warmth|&
            $5.48 \pm 1.08$ &
            $5.60 \pm 0.50$ (0.50, 0.62) &
            $5.84 \pm 0.37$ (1.57, 0.12) &
            $5.72 \pm 0.61$ (0.96, 0.34) \\
        \verb|attuned| &
            $4.00 \pm 1.15$ &
            $4.36 \pm 0.49$ (1.44, 0.16) &
            $4.48 \pm 0.51$ (1.90, 0.06) &
            $4.32 \pm 0.63$ (1.22, 0.23) \\
        \verb|cognitive| &
            $4.76 \pm 0.88$ &
            $5.16 \pm 0.47$ (2.00, 0.05) &
            $5.08 \pm 0.57$ (1.53, 0.13) &
            $5.04 \pm 0.45$ (1.41, 0.16) \\
        \verb|understanding| &
            $4.12 \pm 1.20$ &
            $4.40 \pm 0.50$ (1.08, 0.29) &
            $4.56 \pm 0.51$ (1.69, 0.10) &
            $4.44 \pm 0.51$ (1.23, 0.23) \\
        \verb|acceptance| &
            $5.16 \pm 1.18$ &
            $5.44 \pm 0.51$ (1.09, 0.28) &
            $5.56 \pm 0.51$ (1.56, 0.13) &
            $5.56 \pm 0.51$ (1.56, 0.13) \\
        \midrule
        \verb|average| & 
            $4.70 \pm 1.10$ & 
            $4.99 \pm 0.50$ & 
            $5.07 \pm 0.51$ & 
            $4.97 \pm 0.53$ \\
        \bottomrule
        
    \end{tabular}
    
    \caption{
        Machine (GPT-4o) rated perceived empathy scores (TES item ratings) for each response generation source.
        The mean $\pm$ standard deviation and $t$-test results are shown between machine-generated responses and (professional) human-written responses ($t$-value and $p$-value in brackets).
        Significant differences ($p<0.05$) are bolded.
    }
    \label{tab:eval-mean-sd-gpt-4o}
\end{table*}

\section{System Evaluation}
\label{sec:system-evaluation}

This section describes the research methodology, study design, and corresponding empirical results and discusses them in the context of our system evaluation.
The proposed multi-agent system poses difficulty for a full system evaluation due to the absence of a clear performance benchmark and the presence of multiple functionalities that serve various purposes~\citep{jin2023psyeval, Qiu2023ABF, Rashkin2018TowardsEO}.
Moreover, we present a general-purpose system, which requires further evaluation in its specific application.
In this evaluation, we focus on evaluating the \emph{empathy} of the "Propose response" functionality, a key value proposition in helping mental health professionals and key for therapeutic effectiveness.

\begin{table*}[!ht]

    \centering
    \small

    \begin{tabular}{l | cccc}

        \toprule
        & \multicolumn{4}{c}{Response Generation Source} \\
        \midrule
        TES facet & Human & GPT-4o & Llama 3 70B & Llama 3 8B \\
        \midrule
        \verb|concern| &
            2.33 (\textbf{0.029}) &
            4.39 (\textbf{0.000}) &
            1.59 (0.126) &
            2.53 (\textbf{0.018}) \\
        \verb|resonate| &
            -0.48 (0.637) &
            1.11 (0.278) &
            -1.87 (0.073) &
            -3.59 (\textbf{0.001}) \\
        \verb|warmth|&
            2.75 (\textbf{0.011}) &
            5.06 (\textbf{0.000}) &
            2.49 (\textbf{0.020}) &
            1.55 (0.134) \\
        \verb|attuned| &
            -0.24 (0.811) &
            1.14 (0.265) &
            -1.90 (0.070) &
            -1.63 (0.117) \\
        \verb|cognitive| &
            1.19 (0.246) &
            5.28 (\textbf{0.000}) &
            1.33 (0.196) &
            0.19 (0.852) \\
        \verb|understanding| &
            -0.90 (0.376) &
            1.80 (0.085) &
            -0.58 (0.569) &
            -1.90 (0.070) \\
        \verb|acceptance| &
            0.91 (0.372) &
            3.72 (\textbf{0.001}) &
            0.90 (0.376) &
            0.44 (0.664) \\
        \bottomrule
        
    \end{tabular}
    
    \caption{
        Comparison between human and machine (GPT-4o) rating.
        Paired $t$-test of empathy scores (TES item ratings) between ratings by GPT-4o ($r_1$) and ratings by human raters ($r_2$). 
        Shown are the corresponding $t$-statistics and $p$-values: negative $t$-statistic values indicate that human raters perceived the responses as more empathetic.
        Significant differences ($p<0.05$) are bolded.
    }
    \label{tab:eval-diff-paired-t}
\end{table*}

\subsection{Study Design}

We randomly sampled 100 query-response pairs from the local, anonymous mental health forum previously discussed.
The responses to those queries were written by professional therapists.
To evaluate the quality of responses from LLMs, we used a state-of-the-art commercial model, GPT-4o~\citep{achiam2023gpt}, and a small and large version of a state-of-the-art open source model, Llama 3 with 8 billion and 70 billion parameters, to propose responses for 25 queries each, randomly sampled from the 100 queries, alongside 25 responses from the human therapists.
These responses were then blindly scored against an adaptation of the Therapist Empathy Scale (TES)~\citep{Decker2013TES}, which consists of nine rating items: "concern", "expressiveness" (unused), "\emph{resonate} or capture client feelings", "warmth", "\emph{attuned} to client’s inner world", "understanding \emph{cognitive} framework", "\emph{understanding} feelings/inner experience", "\emph{acceptance} of feelings/inner experiences", and "responsiveness" (unused).
This evaluation excludes the items ``expressiveness'' and ``responsiveness'' because they are largely irrelevant to non-verbal exchanges.
Each item is rated on a 7-point Likert scale based on the extent to which the item is demonstrated in the response, from 1 (not at all) to 7 (extremely).
This adapted version of the TES can be found in Table~\ref{tab:tes-rubrics}.

We consider human evaluation the ground truth, but we are also interested in investigating whether machines can evaluate responses, which would allow for scaling up evaluation efforts.
Therefore, responses were evaluated both by human raters and by GPT-4o.
We recruited ten human raters (three males and seven females), ages 21 to 30.
Each human rater was randomly assigned 10 different query-response pairs from the 100 randomly sampled query-response pairs and tasked with evaluating each response according to the TES.
All human raters had existing mental health or psychological first aid-related training, with some being experienced social workers or counselors-in-training.
Ethical considerations were addressed by briefing all human raters on the nature and purpose of the research and the risks of generative AI.
In addition to human evaluation, we prompted GPT-4o to score each response using the adapted TES.
This was achieved through prompt engineering, where the LLM was tasked to return a score for each facet.

\subsection{Results and Discussion}

Table~\ref{tab:eval-mean-sd-human-raters} shows the means and standard deviations of the human-rated TES scores for each response-generating source.
The $p$-values derived from $t$-tests on the rating scores between the human-generated responses and each of the machine-generated responses are included.
The rating score distributions are shown in Figure~\ref{fig:eval-histogramsHuman}.
Few significant differences are found between the human- and machine-generated responses.
In these evaluations by human raters, Llama 3 8B performed significantly better in the item \verb|resonate| ($p=0.02$), while Llama 3 70B performed significantly better in the item \verb|attuned| ($p=0.04$) as compared to respective facets of empathy in human professional written responses.
This highlights an LLM's comparable ability to capture emotion, thought, tone, and nuance underlying a statement or query, and respond to an individual in distress in a manner that signals active listening and care. 

GPT-4o was also used as a source of evaluation to validate our findings and explore alternative and more scalable evaluation methods.
Table~\ref{tab:eval-mean-sd-gpt-4o} shows the means and standard deviations of the GPT-4o-rated TES rating scores for each response source, as well as the $p$-values derived from $t$-tests on the rating scores between the human-generated responses and each of the machine-generated responses.
The rating score distributions are shown in Figure~\ref{fig:eval-histogramsGPT}.
Again, we do not find many significant differences between human- and machine-generated responses.
In these evaluations by GPT-4o, we only found that responses from GPT-4o performed significantly better in the item \verb|resonate| ($p=0.02$).
The consistently lower standard deviation among LLM responses may be due to the standardized format of LLM responses in demonstrating empathy and the fixed preference of GPT-4o as the sole model for LLM evaluation.
Results from GPT-4o evaluation roughly resonate with results from human evaluation.

In both the AI and human evaluations, LLM responses generally had similar TES ratings to professional responses. 
LLMs only showed a significant improvement in a few rating items of the TES but lacked any significant overall difference when compared to professional responses.
This highlights the overall competency of LLMs in demonstrating a level of empathy comparable to that of a human therapist.
Our results align with previous research on an LLM's capacity for empathy in a therapeutic context.
In the specific context of relationship issues, \citet{Vowels2024helpfulness} found that when evaluating empathy in responses from GPT-4 and relationship experts, results from a sample of 20 human evaluators showed that LLMs performed similarly to or even better than relationship experts.
We stress that our goal is not to show that these LLMs are better than professional therapists, as achieving therapeutic outcomes goes greatly beyond the ability to demonstrate empathy~\citep{Browne-etal-commonfactors}.
Rather, we show that LLMs can propose responses of similar empathetic quality and may be used to augment human-provided care.

\subsection{Comparing LLM and Human Evaluation}

The current reliance on human evaluation is time-consuming and does not scale well, especially since underlying LLMs may be frequently updated.
Reliable LLM evaluation may automate the assessing the domain of empathy.
To determine the reliability of LLM evaluation, we explored the similarity between results from LLM and human evaluations.

In human evaluation (Table~\ref{tab:eval-mean-sd-human-raters}), GPT-4o-generated responses showed the lowest \verb|average| mean ($M = 3.97$), but in LLM evaluation (Table~\ref{tab:eval-mean-sd-gpt-4o}), GPT-4o-generated responses reflected an \verb|average| mean ($M = 4.99$) higher than professional responses. 
One explanation for the disparity between GPT-4o's performance between human and GPT-4o evaluations may be due to the presence of a dataset or AI bias~\citep{srinivasan2021biases}.
When using an LLM model to evaluate a response generated by the same LLM model, the output from both the response and the evaluation comes from the same model trained in the same manner and on the same dataset.
This likely increases the evaluation model's preference towards responses generated by the same model due to the better alignment to the data it was trained on.
Therefore, this results in a higher rating in GPT-4o responses when rated by GPT-4o.
In addition, a common feedback that the AI responses were easily identifiable was heard from a few human evaluators after their evaluations.
This is likely due to the distinct point-form format that GPT-4o uses compared to Llama 3 responses, which use a more elaborative and lengthy format that lacks the point-form structure when responding to most mental health queries.
Given that awareness towards the use of AI use in mental health may lead to prejudice or bias against the AI response~\citep{Jain2024Revealing}, human evaluators may have given the easily identified AI response (GPT-4o responses) a lower rating as compared to responses that had less resemblance to an AI response (human or Llama responses).

A second observed difference between human and LLM evaluation lies in the difference in the reported standard deviations.
Across all response sources, the overall standard deviation in human evaluation was higher than the overall standard deviation in GPT-4o-generated evaluation.
Compared to the GPT-4o evaluation, this increased inconsistency among human raters is likely due to the subjectivity of human interpretations and preferences towards psychological interventions~\citep{Williams-etal-2016-preference}.
Just as how no single therapist can suit every client, each individual has slight preferences.

Table~\ref{tab:eval-diff-paired-t} shows the statistical test results between human and GPT-4o evaluation for each of the four response generation methods.
The GPT-4o evaluation found human responses relatively more empathetic in eight facets across the generation sources, but particularly finds GPT-4o responses more empathetic than humans judged them to be.
These results highlight the differences between LLM and human evaluation.
LLMs may not be suitable for drop-in replacement of human evaluation efforts.
However, the reliability of LLM evaluation and its adherence to human preferences may be further improved through prompt engineering or fine-tuning.

\section{Conclusion}

We presented a general-purpose dual dialogue system to support mental health care providers in their conversations with clients.
The system is designed to assist in crafting responses, recommending resources, and addressing key emotional struggles of care seekers.
It can also reduce the cognitive load and fatigue of care providers.
This system does not aim to replace mental health professionals or promote a thoughtless regurgitation of LLM-generated responses to care seekers.
The human-in-the-loop design allows professional therapists to incorporate the uniqueness of their personality, personal qualities, and style as a therapist~\citep{Smith2003POTT}, and ensure the quality and safety of responses with their professional judgment.
Our evaluation study demonstrated that LLMs are comparable to professional therapists in generating empathetic responses.

\section*{Limitations}

As a system demonstration, we acknowledge several limitations of both the current form of our system and its future applicability. 

First, using a commercial API such as GPT-4o raises privacy concerns, so this system cannot be easily expanded to private conversations.
We intend to explore the use of open-source models such as Llama 3~\citep{Touvron2023Llama2O} and evaluate how this changes the system's performance and acceptability.
Our evaluation study indicates that Llama 3 may be a viable alternative to GPT-4o.

Second, challenges in LLM applications include interpretability, hallucinations~\citep{Ji2023RethinkingLL}, privacy, bias, and clinical effectiveness~\citep{Chung2023ChallengesOL}.
Hallucinations can be obvious, such as nonsensical language or switching to another language, or more subtle, such as returning seemingly correct but erroneous information.
LLMs have also been shown to perpetuate stereotypes of mental disorders such as schizophrenia~\citep{King2022HarmfulBI}.

Third, engagement remains an issue in this space.
We already know that mental health apps fail to consistently engage users.
However, tools for clinicians are often not engaging either.
For the current system to be embraced and adopted, we intend to conduct further usability studies~\citep{Greenhalgh2017BeyondAA}.

Fourth, while we have specifically framed this system as a human-in-the-loop system, we acknowledge that, in practice, this distinction may fade.
When the proposed responses are good, and care professionals are overwhelmed, the proposed responses may not be validated by humans anymore as an unintended consequence (an effect known as \emph{automation bias}).
This reintroduces some of the risks of a system that directly interacts with clients.
This phenomenon has been observed in self-driving cars, where drivers start to trust the system and stop paying attention to such a degree that they can no longer be considered a robust safety agent in the loop.

Fifth, our results' generalizability is limited by our small sample size of 100 query responses and ten human evaluators.
The human evaluators, all between 21 and 30 and 70 percent female, may skew our results.
Furthermore, the sole focus on empathy will not be sufficient to test the utility of the system at large.
Therefore, an evaluation with a larger sample to ensure the generalizability of our results and a system-as-a-whole evaluation is needed.

\section*{Ethics Statement}

The dual dialogue system discussed may have far-reaching societal impacts.
It may represent, at least in part, a redeployment of resources towards technology that could otherwise have been used to train more mental health professionals.
Moreover, rapid proliferation of such systems could destabilize professional norms and institutional practices in unsustainable ways~\citep{wang2024human}.
However, we reiterate that our system is designed to \textit{support} therapists, helping them to better help their clients rather than to replace them.

Current state-of-the-art LLMs exhibit markedly stronger performance in English than in other languages.
In diverse and multilingual societies such as Singapore, a system that only works well in English is likely to be inadequate for many users.
It may even be the case that, given the nature of existing service provision, it is even more important to serve users in languages other than English~\citep{lovenia2024seacrowd}.
We hope that future evaluations can explore system performance in other languages in greater detail and consider the evaluation of models explicitly prioritizing multilinguality~\citep{ustun2024aya}. 

LLMs are not value-neutral and have been shown to encode specific social, cultural, and political viewpoints, often aligned with North American values~\citep{santurkarWhoseOpinionsLanguage2023, aroraProbingPreTrainedLanguage2023}.
Mental health conditions and the seeking of mental health support are often stigmatized in various ways around the world. 
For an LLM-based mental health tool to be effective, it may need to understand the specific socio-cultural context of its users, which may engender additional fine-tuning by incorporating diverse and representative viewpoints.

As this work is a demonstration of an early system, it has not yet been extensively evaluated with real human participants in (one of) its applications, which includes the system operators and the downstream end users.
All conversation extracts used in our analysis and included in the figures throughout this work were taken from public forum responses.
Evaluating the performance and acceptability of this system with real users seeking mental health support would require a thorough ethics review by an institutional review board. 

\section*{Acknowledgments}

We thank Samuel J. Bell for his helpful feedback on our manuscript.
We also thank our human raters: Kellie Sim, Lee Li Hwee, Zuyuan Cheong, and Allyson Grace Chai Y. J. for their contributions.
We thank Claire Seah for diagram design.

\bibliography{custom}

\begin{thebibliography}{71}
\providecommand{\natexlab}[1]{#1}

\bibitem[{Achiam et~al.(2023)Achiam, Adler, Agarwal, Ahmad, Akkaya, Aleman, Almeida, Altenschmidt, Altman, Anadkat et~al.}]{achiam2023gpt}
Josh Achiam, Steven Adler, Sandhini Agarwal, Lama Ahmad, Ilge Akkaya, Florencia~Leoni Aleman, Diogo Almeida, Janko Altenschmidt, Sam Altman, Shyamal Anadkat, et~al. 2023.
\newblock {GPT-4 technical report}.
\newblock \emph{arXiv preprint arXiv:2303.08774}.

\bibitem[{Althoff et~al.(2016)Althoff, Clark, and Leskovec}]{althoff2016large}
Tim Althoff, Kevin Clark, and Jure Leskovec. 2016.
\newblock Large-scale analysis of counseling conversations: An application of natural language processing to mental health.
\newblock \emph{Transactions of the Association for Computational Linguistics}, 4:463--476.

\bibitem[{Amin et~al.(2023)Amin, Cambria, and Schuller}]{amin_will_2023}
Mostafa~M. Amin, Erik Cambria, and Björn~W. Schuller. 2023.
\newblock \href {http://arxiv.org/abs/2303.03186} {Will {Affective} {Computing} {Emerge} from {Foundation} {Models} and {General} {AI}? {A} {First} {Evaluation} on {ChatGPT}}.
\newblock \emph{arXiv preprint}.

\bibitem[{Arora et~al.(2023)Arora, Kaffee, and Augenstein}]{aroraProbingPreTrainedLanguage2023}
Arnav Arora, Lucie-aim{\'e}e Kaffee, and Isabelle Augenstein. 2023.
\newblock \href {https://doi.org/10.18653/v1/2023.c3nlp-1.12} {Probing {{Pre-Trained Language Models}} for {{Cross-Cultural Differences}} in {{Values}}}.
\newblock In \emph{Proceedings of the {{First Workshop}} on {{Cross-Cultural Considerations}} in {{NLP}} ({{C3NLP}})}, pages 114--130.

\bibitem[{Bendig et~al.(2019)Bendig, Erb, Schulze-Thuesing, and Baumeister}]{Bendig2019TheNG}
Eileen Bendig, Benjamin Erb, Lea Schulze-Thuesing, and Harald Baumeister. 2019.
\newblock \href {https://api.semanticscholar.org/CorpusID:204394372} {The next generation: Chatbots in clinical psychology and psychotherapy to foster mental health – a scoping review}.
\newblock \emph{Verhaltenstherapie}, 32:64 -- 76.

\bibitem[{Brocki et~al.(2023)Brocki, Dyer, Gładka, and Chung}]{Brocki2023DeepLM}
Lennart Brocki, George~C. Dyer, Anna Gładka, and Neo~Christopher Chung. 2023.
\newblock \href {https://api.semanticscholar.org/CorpusID:256105728} {Deep learning mental health dialogue system}.
\newblock \emph{2023 IEEE International Conference on Big Data and Smart Computing (BigComp)}, pages 395--398.

\bibitem[{Browne et~al.(2021)Browne, Cather, and Mueser}]{Browne-etal-commonfactors}
Julia Browne, Corinne Cather, and Kim~T. Mueser. 2021.
\newblock \href {https://doi.org/10.1093/acrefore/9780190236557.013.79} {Common factors in psychotherapy}.

\bibitem[{{Caceres Najarro} et~al.(2023){Caceres Najarro}, Lee, Toshnazarov, Jang, Kim, and Noh}]{CaceresNajarro2023WMGPTT2}
Lismer~Andres {Caceres Najarro}, Yonggeon Lee, Kobiljon~E. Toshnazarov, Yoonhyung Jang, Hyungsook Kim, and Youngtae Noh. 2023.
\newblock \href {https://api.semanticscholar.org/CorpusID:263742958} {{WMGPT: Towards 24/7 online prime counseling with ChatGPT}}.
\newblock \emph{Adjunct Proceedings of the 2023 ACM International Joint Conference on Pervasive and Ubiquitous Computing \& the 2023 ACM International Symposium on Wearable Computing}.

\bibitem[{Cameron et~al.(2017)Cameron, Cameron, Megaw, Bond, Mulvenna, O’neill, Armour, and McTear}]{Cameron2017TowardsAC}
Gillian Cameron, David Cameron, Gavin Megaw, Raymond~R. Bond, Maurice~D. Mulvenna, Siobhan O’neill, Ch{\'e}rie Armour, and Michael~F. McTear. 2017.
\newblock \href {https://api.semanticscholar.org/CorpusID:4796070} {Towards a chatbot for digital counselling}.
\newblock In \emph{British Computer Society Conference on Human-Computer Interaction}.

\bibitem[{Chase(2022)}]{Chase_LangChain_2022}
Harrison Chase. 2022.
\newblock \href {https://github.com/langchain-ai/langchain} {{LangChain}}.

\bibitem[{Chung et~al.(2023)Chung, Dyer, and Brocki}]{Chung2023ChallengesOL}
Neo~Christopher Chung, George~C. Dyer, and Lennart Brocki. 2023.
\newblock \href {https://api.semanticscholar.org/CorpusID:265445592} {Challenges of large language models for mental health counseling}.
\newblock \emph{ArXiv}, abs/2311.13857.

\bibitem[{Decker et~al.(2013)Decker, Nich, Carroll, and Martino}]{Decker2013TES}
Suzanne~E. Decker, Charla Nich, Kathleen~M. Carroll, and Steve Martino. 2013.
\newblock \href {https://doi.org/10.1017/s1352465813000039} {Development of the therapist empathy scale}.
\newblock \emph{Behavioural and Cognitive Psychotherapy}, 42:339 -- 354.

\bibitem[{Deepset(2019)}]{Deepset_Haystack_2019}
Deepset. 2019.
\newblock \href {https://github.com/deepset-ai/haystack} {{Haystack}}.

\bibitem[{Douze et~al.(2024)Douze, Guzhva, Deng, Johnson, Szilvasy, Mazar{\'e}, Lomeli, Hosseini, and J{\'e}gou}]{douze2024faiss}
Matthijs Douze, Alexandr Guzhva, Chengqi Deng, Jeff Johnson, Gergely Szilvasy, Pierre-Emmanuel Mazar{\'e}, Maria Lomeli, Lucas Hosseini, and Herv{\'e} J{\'e}gou. 2024.
\newblock The faiss library.
\newblock \emph{arXiv preprint arXiv:2401.08281}.

\bibitem[{Fitzpatrick et~al.(2017)Fitzpatrick, Darcy, and Vierhile}]{Fitzpatrick2017DeliveringCB}
Kathleen~Kara Fitzpatrick, Alison~M Darcy, and Molly Vierhile. 2017.
\newblock \href {https://api.semanticscholar.org/CorpusID:3772810} {{Delivering Cognitive Behavior Therapy to Young Adults With Symptoms of Depression and Anxiety Using a Fully Automated Conversational Agent (Woebot): A Randomized Controlled Trial}}.
\newblock \emph{JMIR Mental Health}, 4.

\bibitem[{Fung et~al.(2016)Fung, Bertero, Wan, Dey, Chan, Siddique, Yang, Wu, and Lin}]{Fung2016TowardsEH}
Pascale Fung, Dario Bertero, Yan Wan, Anik Dey, Ricky Ho~Yin Chan, Farhad~Bin Siddique, Yang Yang, Chien-Sheng Wu, and Ruixi Lin. 2016.
\newblock \href {https://api.semanticscholar.org/CorpusID:4056254} {Towards empathetic human-robot interactions}.
\newblock In \emph{Conference on Intelligent Text Processing and Computational Linguistics}.

\bibitem[{Greenhalgh et~al.(2017)Greenhalgh, Wherton, Papoutsi, Lynch, Hughes, A'Court, Hinder, Fahy, Procter, and Shaw}]{Greenhalgh2017BeyondAA}
Trisha Greenhalgh, Joseph~P. Wherton, Chrysanthi Papoutsi, Jennifer Lynch, Gemma Hughes, Christine A'Court, Susan Hinder, Nick Fahy, R.~N. Procter, and Sara~E. Shaw. 2017.
\newblock \href {https://api.semanticscholar.org/CorpusID:3818961} {Beyond adoption: A new framework for theorizing and evaluating nonadoption, abandonment, and challenges to the scale-up, spread, and sustainability of health and care technologies}.
\newblock \emph{Journal of Medical Internet Research}, 19.

\bibitem[{Griffiths et~al.(2012)Griffiths, Mackinnon, Crisp, Christensen, Bennett, and Farrer}]{Griffiths2012TheEO}
Kathleen~Margaret Griffiths, Andrew~J Mackinnon, Dimity~Ann Crisp, Helen Christensen, Kylie Bennett, and Louise~M. Farrer. 2012.
\newblock \href {https://api.semanticscholar.org/CorpusID:263944624} {The effectiveness of an online support group for members of the community with depression: A randomised controlled trial}.
\newblock \emph{PLoS ONE}, 7.

\bibitem[{Hua et~al.(2024)Hua, Liu, Yang, Li, Sheu, Zhou, Moran, Ananiadou, and Beam}]{hua2024large}
Yining Hua, Fenglin Liu, Kailai Yang, Zehan Li, Yi-han Sheu, Peilin Zhou, Lauren~V Moran, Sophia Ananiadou, and Andrew Beam. 2024.
\newblock {Large Language Models in Mental Health Care: A Scoping Review}.
\newblock \emph{arXiv preprint arXiv:2401.02984}.

\bibitem[{Inkster et~al.(2018)Inkster, Sarda, and Subramanian}]{Inkster2018AnEC}
Becky Inkster, Shubhankar Sarda, and Vinod Subramanian. 2018.
\newblock \href {https://api.semanticscholar.org/CorpusID:53719693} {{An Empathy-Driven, Conversational Artificial Intelligence Agent (Wysa) for Digital Mental Well-Being: Real-World Data Evaluation Mixed-Methods Study}}.
\newblock \emph{JMIR mHealth and uHealth}, 6.

\bibitem[{Irarrázaval and Kalawski(2022)}]{Irarrazaval2022Empathy}
Leonor Irarrázaval and Juan~Pablo Kalawski. 2022.
\newblock \href {https://doi.org/10.3389/fpsyg.2022.1000059} {Phenomenological considerations on empathy and emotions in psychotherapy}.
\newblock \emph{Frontiers in Psychology}, 13.

\bibitem[{Jain et~al.(2024)Jain, Pareek, and Carlbring}]{Jain2024Revealing}
Gagan Jain, Samridhi Pareek, and Per Carlbring. 2024.
\newblock \href {https://doi.org/10.1016/j.invent.2024.100745} {{Revealing the source: How awareness alters perceptions of AI and human-generated mental health responses}}.
\newblock \emph{Internet Interventions}, 36:100745.

\bibitem[{Ji et~al.(2023{\natexlab{a}})Ji, Zhang, Yang, Ananiadou, and Cambria}]{Ji2023RethinkingLL}
Shaoxiong Ji, Tianlin Zhang, Kailai Yang, Sophia Ananiadou, and Erik Cambria. 2023{\natexlab{a}}.
\newblock \href {https://api.semanticscholar.org/CorpusID:265295232} {Rethinking large language models in mental health applications}.
\newblock \emph{ArXiv}, abs/2311.11267.

\bibitem[{Ji et~al.(2023{\natexlab{b}})Ji, Lee, Frieske, Yu, Su, Xu, Ishii, Bang, Madotto, and Fung}]{ji2023survey}
Ziwei Ji, Nayeon Lee, Rita Frieske, Tiezheng Yu, Dan Su, Yan Xu, Etsuko Ishii, Ye~Jin Bang, Andrea Madotto, and Pascale Fung. 2023{\natexlab{b}}.
\newblock Survey of hallucination in natural language generation.
\newblock \emph{ACM Computing Surveys}, 55(12):1--38.

\bibitem[{Jin et~al.(2023)Jin, Chen, Wu, and Zhu}]{jin2023psyeval}
Haoan Jin, Siyuan Chen, Mengyue Wu, and Kenny~Q Zhu. 2023.
\newblock {PsyEval: A Comprehensive Large Language Model Evaluation Benchmark for Mental Health}.
\newblock \emph{arXiv preprint arXiv:2311.09189}.

\bibitem[{King(2022)}]{King2022HarmfulBI}
Morgan King. 2022.
\newblock \href {https://api.semanticscholar.org/CorpusID:252917188} {Harmful biases in artificial intelligence.}
\newblock \emph{The Lancet Psychiatry}.

\bibitem[{Kwek et~al.(2024)Kwek, Yim, Andersson, Suendermann, Subramaniam, Yadin, Vaingankar, and Gupta}]{kwek2024effectiveness}
Tammie Rong~Rong Kwek, Jackki Hoon~Eng Yim, Erik Andersson, Oliver Suendermann, Mythily Subramaniam, Elna Yadin, Janhavi Vaingankar, and Bhanu Gupta. 2024.
\newblock {Effectiveness and acceptability of Internet-based Cognitive Behavioral Therapy for individuals with Obsessive Compulsive Disorder in Singapore}.
\newblock \emph{Journal of Behavioral and Cognitive Therapy}, 34(2):100487.

\bibitem[{Lai et~al.(2023)Lai, Shi, Du, Wu, Fu, Dou, and Wang}]{Lai2023PsyLLMSU}
Tin Lai, Yukun Shi, Zicong Du, Jiajie Wu, Ken Fu, Yichao Dou, and Ziqi Wang. 2023.
\newblock \href {https://api.semanticscholar.org/CorpusID:260125719} {{Psy-LLM: Scaling up Global Mental Health Psychological Services with AI-based Large Language Models}}.
\newblock \emph{ArXiv}, abs/2307.11991.

\bibitem[{Lee et~al.(2023)Lee, Goldberg, and Kohane}]{lee2023ai}
Peter Lee, Carey Goldberg, and Isaac Kohane. 2023.
\newblock \emph{The AI revolution in medicine: GPT-4 and beyond}.
\newblock Pearson.

\bibitem[{Lee et~al.(2020)Lee, Yamashita, Huang, and Fu}]{Lee2020IHY}
Yi-Chieh Lee, Naomi Yamashita, Yun Huang, and Wai-Tat Fu. 2020.
\newblock \href {https://api.semanticscholar.org/CorpusID:218482970} {{"I Hear You, I Feel You": Encouraging Deep Self-disclosure through a Chatbot}}.
\newblock \emph{Proceedings of the 2020 CHI Conference on Human Factors in Computing Systems}.

\bibitem[{Li et~al.(2023)Li, Zhang, Lee, Kraut, and Mohr}]{li2023systematic}
Han Li, Renwen Zhang, Yi-Chieh Lee, Robert~E Kraut, and David~C Mohr. 2023.
\newblock {Systematic review and meta-analysis of AI-based conversational agents for promoting mental health and well-being}.
\newblock \emph{NPJ Digital Medicine}, 6(1):236.

\bibitem[{Linardon et~al.(2024)Linardon, Torous, Firth, Cuijpers, Messer, and Fuller-Tyszkiewicz}]{Linardon2024CurrentEO}
Jake Linardon, John~B Torous, Joseph Firth, Pim Cuijpers, Mariel Messer, and Matthew Fuller-Tyszkiewicz. 2024.
\newblock \href {https://api.semanticscholar.org/CorpusID:266962875} {Current evidence on the efficacy of mental health smartphone apps for symptoms of depression and anxiety: A meta‐analysis of 176 randomized controlled trials}.
\newblock \emph{World Psychiatry}, 23.

\bibitem[{Liu(2022)}]{Liu_LlamaIndex_2022}
Jerry Liu. 2022.
\newblock \href {https://github.com/run-llama/llama_index} {{LlamaIndex}}.

\bibitem[{Lovenia et~al.(2024)Lovenia, Mahendra, Akbar, Miranda, Santoso, Aco, Fadhilah, Mansurov, Imperial, Kampman, Moniz, Habibi, Hudi, Montalan, Ignatius, Lopo, Nixon, Karlsson, Jaya, Diandaru, Gao, Amadeus, Wang, Cruz, Whitehouse, Parmonangan, Khelli, Zhang, Susanto, Ryanda, Hermawan, Velasco, Kautsar, Hendria, Moslem, Flynn, Adilazuarda, Li, Lee, Damanhuri, Sun, Qorib, Djanibekov, Leong, Do, Muennighoff, Pansuwan, Putra, Xu, Tai, Purwarianti, Ruder, Tjhi, Limkonchotiwat, Aji, Keh, Winata, Zhang, Koto, Yong, and Cahyawijaya}]{lovenia2024seacrowd}
Holy Lovenia, Rahmad Mahendra, Salsabil~Maulana Akbar, Lester James~V. Miranda, Jennifer Santoso, Elyanah Aco, Akhdan Fadhilah, Jonibek Mansurov, Joseph~Marvin Imperial, Onno~P. Kampman, Joel Ruben~Antony Moniz, Muhammad Ravi~Shulthan Habibi, Frederikus Hudi, Railey Montalan, Ryan Ignatius, Joanito~Agili Lopo, William Nixon, Börje~F. Karlsson, James Jaya, Ryandito Diandaru, Yuze Gao, Patrick Amadeus, Bin Wang, Jan Christian~Blaise Cruz, Chenxi Whitehouse, Ivan~Halim Parmonangan, Maria Khelli, Wenyu Zhang, Lucky Susanto, Reynard~Adha Ryanda, Sonny~Lazuardi Hermawan, Dan~John Velasco, Muhammad Dehan~Al Kautsar, Willy~Fitra Hendria, Yasmin Moslem, Noah Flynn, Muhammad~Farid Adilazuarda, Haochen Li, Johanes Lee, R.~Damanhuri, Shuo Sun, Muhammad~Reza Qorib, Amirbek Djanibekov, Wei~Qi Leong, Quyet~V. Do, Niklas Muennighoff, Tanrada Pansuwan, Ilham~Firdausi Putra, Yan Xu, Ngee~Chia Tai, Ayu Purwarianti, Sebastian Ruder, William Tjhi, Peerat Limkonchotiwat, Alham~Fikri Aji, Sedrick Keh, Genta~Indra Winata, Ruochen
  Zhang, Fajri Koto, Zheng-Xin Yong, and Samuel Cahyawijaya. 2024.
\newblock \href {https://arxiv.org/abs/2406.10118} {{SEACrowd: A Multilingual Multimodal Data Hub and Benchmark Suite for Southeast Asian Languages}}.
\newblock \emph{Proceedings of EMNLP}.

\bibitem[{Lu et~al.(2021)Lu, Assudani, Kwek, Ng, Teoh, and Tan}]{lu2021randomised}
Sharon~HX Lu, Hanita~A Assudani, Tammie~RR Kwek, Shaun~WH Ng, Trisha~EL Teoh, and Geoffrey~CY Tan. 2021.
\newblock A randomised controlled trial of clinician-guided internet-based cognitive behavioural therapy for depressed patients in singapore.
\newblock \emph{Frontiers in Psychology}, 12.

\bibitem[{Luo et~al.(2024)Luo, Warren, Cheng, Abdul-Muhsin, and Banerjee}]{Luo2024empathy}
Man Luo, Christopher~J. Warren, Lu~Cheng, Haidar~M. Abdul-Muhsin, and Imon Banerjee. 2024.
\newblock \href {https://doi.org/10.48550/arXiv.2405.16402} {Assessing empathy in large language models with real-world physician-patient interactions}.
\newblock \emph{arXiv preprint}.

\bibitem[{Ma et~al.(2020)Ma, Nguyen, Xing, and Cambria}]{Ma2020ASO}
Yukun Ma, Khanh~Linh Nguyen, Frank Xing, and E.~Cambria. 2020.
\newblock \href {https://api.semanticscholar.org/CorpusID:219178913} {A survey on empathetic dialogue systems}.
\newblock \emph{Inf. Fusion}, 64:50--70.

\bibitem[{Maddela et~al.(2023)Maddela, Ung, Xu, Madotto, Foran, and Boureau}]{maddela2023training}
Mounica Maddela, Megan Ung, Jing Xu, Andrea Madotto, Heather Foran, and Y-Lan Boureau. 2023.
\newblock Training models to generate, recognize, and reframe unhelpful thoughts.
\newblock \emph{arXiv preprint arXiv:2307.02768}.

\bibitem[{Malgaroli et~al.(2023)Malgaroli, Hull, Zech, and Althoff}]{Malgaroli2023NaturalLP}
Matteo Malgaroli, Thomas~D. Hull, Jamie~M. Zech, and Tim Althoff. 2023.
\newblock \href {https://api.semanticscholar.org/CorpusID:263703571} {Natural language processing for mental health interventions: A systematic review and research framework}.
\newblock \emph{Translational Psychiatry}, 13.

\bibitem[{{Ministry of Health Singapore}(2023)}]{MOHstats}
{Ministry of Health Singapore}. 2023.
\newblock Waiting time for appointments for mental health treatment at public healthcare institutions.
\newblock https://www.moh.gov.sg/news-highlights/details/waiting-time-for-appointments-for-mental-health-treatment-at-public-healthcare-institutions.

\bibitem[{O'Connor et~al.(2018)O'Connor, Muller~Neff, and Pitman}]{oconnor2018burnout}
Karen O'Connor, Deirdre Muller~Neff, and Steve Pitman. 2018.
\newblock \href {https://doi.org/10.1016/j.eurpsy.2018.06.003} {Burnout in mental health professionals: A systematic review and meta-analysis of prevalence and determinants}.
\newblock \emph{European psychiatry: The journal of the Association of European Psychiatrists}, 53:74--99.

\bibitem[{Phang et~al.(2023)Phang, Heaukulani, Martanto, Morris, Tong, and Ho}]{phang2023perceptions}
Ye~Sheng Phang, Creighton Heaukulani, Wijaya Martanto, Robert Morris, Mian~Mian Tong, and Roger Ho. 2023.
\newblock \href {https://doi.org/10.2196/42167} {{Perceptions of a Digital Mental Health Platform Among Participants With Depressive Disorder, Anxiety Disorder, and Other Clinically Diagnosed Mental Disorders in Singapore: Usability and Acceptability Study}}.
\newblock \emph{JMIR Human Factors}, 10:e42167.

\bibitem[{Priyadarshana et~al.(2024)Priyadarshana, Senanayake, Liang, and Piumarta}]{priyadarshana6prompt}
Y.H.P.P. Priyadarshana, Ashala Senanayake, Zilu Liang, and Ian Piumarta. 2024.
\newblock Prompt engineering for digital mental health: A short review.
\newblock \emph{Frontiers in Digital Health}, 6:1410947.

\bibitem[{Qiu et~al.(2023)Qiu, Zhao, Li, Zhang, He, and Lan}]{Qiu2023ABF}
Huachuan Qiu, Tong Zhao, Anqi Li, Shuai Zhang, Hongliang He, and Zhenzhong Lan. 2023.
\newblock \href {https://api.semanticscholar.org/CorpusID:260334700} {A benchmark for understanding dialogue safety in mental health support}.
\newblock In \emph{Natural Language Processing and Chinese Computing}.

\bibitem[{Rashkin et~al.(2018)Rashkin, Smith, Li, and Boureau}]{Rashkin2018TowardsEO}
Hannah Rashkin, Eric~Michael Smith, Margaret Li, and Y-Lan Boureau. 2018.
\newblock \href {https://api.semanticscholar.org/CorpusID:195069365} {Towards empathetic open-domain conversation models: A new benchmark and dataset}.
\newblock In \emph{Annual Meeting of the Association for Computational Linguistics}.

\bibitem[{Rogers(1957)}]{Rogers1957}
Carl~R. Rogers. 1957.
\newblock \href {https://doi.org/10.1037/h0045357} {The necessary and sufficient conditions of therapeutic personality change}.
\newblock \emph{Journal of Consulting Psychology}, 21:95–103.

\bibitem[{Santurkar et~al.(2023)Santurkar, Durmus, Ladhak, Lee, Liang, and Hashimoto}]{santurkarWhoseOpinionsLanguage2023}
Shibani Santurkar, Esin Durmus, Faisal Ladhak, Cinoo Lee, Percy Liang, and Tatsunori Hashimoto. 2023.
\newblock \href {https://doi.org/10.48550/arXiv.2303.17548} {Whose {{Opinions Do Language Models Reflect}}?}
\newblock \emph{Preprint}, arxiv:2303.17548.

\bibitem[{Shanahan et~al.(2023)Shanahan, McDonell, and Reynolds}]{Shanahan2023RolePW}
Murray Shanahan, Kyle McDonell, and Laria Reynolds. 2023.
\newblock \href {https://api.semanticscholar.org/CorpusID:258947657} {Role play with large language models}.
\newblock \emph{Nature}, 623:493--498.

\bibitem[{{Singapore Medical Council}(2023)}]{SingaporeMedicalCouncil2023}
{Singapore Medical Council}. 2023.
\newblock Annual report 2023.
\newblock \url{https://www.healthprofessionals.gov.sg/docs/librariesprovider2/publications-newsroom/smc-annual-reports/smc_annual-report-2023-03sept-pdfa-(1).pdf?sfvrsn=b379c36a_1}.

\bibitem[{{Singapore Psychological Society}(2024)}]{SingaporePsychologicalSociety}
{Singapore Psychological Society}. 2024.
\newblock Directory of sps members.
\newblock \url{https://singaporepsychologicalsociety.org/members-directory/}.

\bibitem[{Smith(2003)}]{Smith2003POTT}
Edward~WL Smith. 2003.
\newblock \emph{The person of the therapist}.
\newblock McFarland.

\bibitem[{Srinivasan and Chander(2021)}]{srinivasan2021biases}
Ramya Srinivasan and Ajay Chander. 2021.
\newblock Biases in ai systems.
\newblock \emph{Communications of the ACM}, 64(8):44--49.

\bibitem[{Subramaniam et~al.(2019)Subramaniam, Abdin, Vaingankar, Shafie, Chua, Sambasivam, Zhang, Shahwan, Chang, Chua, Verma, James, Kwok, Heng, and Chong}]{Subramaniam2019TrackingTM}
Mythily Subramaniam, Edimansyah Abdin, Janhavi~Ajit Vaingankar, Saleha Shafie, Boon~Yiang Chua, Rajeswari Sambasivam, Y.~J. Zhang, Shazana Shahwan, S.~M. Chang, Hong~Choon Chua, Swapna Verma, Lyn James, Kian~Woon Kwok, Derrick~Mk Heng, and Siow~Ann Chong. 2019.
\newblock \href {https://api.semanticscholar.org/CorpusID:96434454} {{Tracking the mental health of a nation: Prevalence and correlates of mental disorders in the second Singapore mental health study}}.
\newblock \emph{Epidemiology and Psychiatric Sciences}, 29.

\bibitem[{Torous and Blease(2024)}]{Torous2024GenerativeAI}
John~B. Torous and Charlotte Blease. 2024.
\newblock \href {https://api.semanticscholar.org/CorpusID:266960503} {Generative artificial intelligence in mental health care: Potential benefits and current challenges}.
\newblock \emph{World Psychiatry}, 23.

\bibitem[{Touvron et~al.(2023)Touvron, Martin, Stone, Albert, Almahairi, Babaei, Bashlykov, Batra, Bhargava, Bhosale, Bikel, Blecher, Ferrer, Chen, Cucurull, Esiobu, Fernandes, Fu, Fu, Fuller, Gao, Goswami, Goyal, Hartshorn, Hosseini, Hou, Inan, Kardas, Kerkez, Khabsa, Kloumann, Korenev, Koura, Lachaux, Lavril, Lee, Liskovich, Lu, Mao, Martinet, Mihaylov, Mishra, Molybog, Nie, Poulton, Reizenstein, Rungta, Saladi, Schelten, Silva, Smith, Subramanian, Tan, Tang, Taylor, Williams, Kuan, Xu, Yan, Zarov, Zhang, Fan, Kambadur, Narang, Rodriguez, Stojnic, Edunov, and Scialom}]{Touvron2023Llama2O}
Hugo Touvron, Louis Martin, Kevin~R. Stone, Peter Albert, Amjad Almahairi, Yasmine Babaei, Nikolay Bashlykov, Soumya Batra, Prajjwal Bhargava, Shruti Bhosale, Daniel~M. Bikel, Lukas Blecher, Cristian~Cant{\'o}n Ferrer, Moya Chen, Guillem Cucurull, David Esiobu, Jude Fernandes, Jeremy Fu, Wenyin Fu, Brian Fuller, Cynthia Gao, Vedanuj Goswami, Naman Goyal, Anthony~S. Hartshorn, Saghar Hosseini, Rui Hou, Hakan Inan, Marcin Kardas, Viktor Kerkez, Madian Khabsa, Isabel~M. Kloumann, A.~V. Korenev, Punit~Singh Koura, Marie-Anne Lachaux, Thibaut Lavril, Jenya Lee, Diana Liskovich, Yinghai Lu, Yuning Mao, Xavier Martinet, Todor Mihaylov, Pushkar Mishra, Igor Molybog, Yixin Nie, Andrew Poulton, Jeremy Reizenstein, Rashi Rungta, Kalyan Saladi, Alan Schelten, Ruan Silva, Eric~Michael Smith, R.~Subramanian, Xia Tan, Binh Tang, Ross Taylor, Adina Williams, Jian~Xiang Kuan, Puxin Xu, Zhengxu Yan, Iliyan Zarov, Yuchen Zhang, Angela Fan, Melanie Kambadur, Sharan Narang, Aurelien Rodriguez, Robert Stojnic, Sergey Edunov, and
  Thomas Scialom. 2023.
\newblock \href {https://api.semanticscholar.org/CorpusID:259950998} {Llama 2: Open foundation and fine-tuned chat models}.
\newblock \emph{ArXiv}, abs/2307.09288.

\bibitem[{{\"U}st{\"u}n et~al.(2024){\"U}st{\"u}n, Aryabumi, Yong, Ko, D'souza, Onilude, Bhandari, Singh, Ooi, Kayid, Vargus, Blunsom, Longpre, Muennighoff, Fadaee, Kreutzer, and Hooker}]{ustun2024aya}
Ahmet {\"U}st{\"u}n, Viraat Aryabumi, Zheng-Xin Yong, Wei-Yin Ko, Daniel D'souza, Gbemileke Onilude, Neel Bhandari, Shivalika Singh, Hui-Lee Ooi, Amr Kayid, Freddie Vargus, Phil Blunsom, Shayne Longpre, Niklas Muennighoff, Marzieh Fadaee, Julia Kreutzer, and Sara Hooker. 2024.
\newblock \href {https://doi.org/10.48550/arXiv.2402.07827} {Aya {{Model}}: {{An Instruction Finetuned Open-Access Multilingual Language Model}}}.
\newblock \emph{Preprint}, arxiv:2402.07827.

\bibitem[{Vaidyam et~al.(2019)Vaidyam, Wisniewski, Halamka, Kashavan, and Torous}]{Vaidyam2019ChatbotsAC}
Aditya~Nrusimha Vaidyam, Hannah Wisniewski, John~D. Halamka, Matcheri~S. Kashavan, and John~B Torous. 2019.
\newblock \href {https://api.semanticscholar.org/CorpusID:85446172} {Chatbots and conversational agents in mental health: A review of the psychiatric landscape}.
\newblock \emph{The Canadian Journal of Psychiatry}, 64:456 -- 464.

\bibitem[{van Heerden et~al.(2023)van Heerden, Pozuelo, and Kohrt}]{vanHeerden2023GlobalMH}
Alastair~C. van Heerden, Julia~R Pozuelo, and Brandon~A. Kohrt. 2023.
\newblock \href {https://api.semanticscholar.org/CorpusID:258743117} {Global mental health services and the impact of artificial intelligence-powered large language models.}
\newblock \emph{JAMA psychiatry}.

\bibitem[{Vowels(2024)}]{Vowels2024helpfulness}
Laura~M. Vowels. 2024.
\newblock \href {https://doi.org/10.1016/j.chbah.2024.100077} {{Are chatbots the new relationship experts? Insights from three studies}}.
\newblock \emph{Computers in Human Behavior: Artificial Humans}, 2.

\bibitem[{Vowels et~al.(2024)Vowels, Francois-Walcott, and Darwiche}]{Vowels2024LLMtherapist}
Laura~M. Vowels, Rachel~R.R. Francois-Walcott, and Joëlle Darwiche. 2024.
\newblock \href {https://doi.org/10.1016/j.chbah.2024.100078} {{AI in Relationship Counselling: Evaluating ChatGPT’s Therapeutic Capabilities in Providing Relationship Advice}}.
\newblock \emph{Computers in Human Behavior: Artificial Humans}, 2.

\bibitem[{Wang et~al.(2024)Wang, Cooper, and Eby}]{wang2024human}
Skyler Wang, Ned Cooper, and Margaret Eby. 2024.
\newblock {From human-centered to social-centered artificial intelligence: Assessing ChatGPT's impact through disruptive events}.
\newblock \emph{Big Data \& Society}, 11(4).

\bibitem[{Weizenbaum(1966)}]{weizenbaum1966eliza}
Joseph Weizenbaum. 1966.
\newblock {ELIZA - A computer program for the study of natural language communication between man and machine}.
\newblock \emph{Communications of the ACM}, 9(1):36--45.

\bibitem[{Weng et~al.(2024)Weng, Hu, Heaukulani, Tan, Chang, Phang, Rajendram, Tan, Loke, and Morris}]{weng2024mental}
Janice~Huiqin Weng, Yanyan Hu, Creighton Heaukulani, Clarence Tan, Julian~Kuiyu Chang, Ye~Sheng Phang, Priyanka Rajendram, Weng~Mooi Tan, Wai~Chiong Loke, and Robert~JT Morris. 2024.
\newblock {Mental Wellness Self-Care in Singapore With mindline.sg: A Tutorial on the Development of a Digital Mental Health Platform for Behavior Change}.
\newblock \emph{Journal of Medical Internet Research}, 26.

\bibitem[{Williams et~al.(2016)Williams, Farquharson, Palmer, Bassett, Clarke, Clark, and Crawford}]{Williams-etal-2016-preference}
Ryan Williams, Lorna Farquharson, Lucy Palmer, Paul Bassett, Jeremy Clarke, David~M. Clark, and Mike~J. Crawford. 2016.
\newblock {Patient preference in psychological treatment and associations with self-reported outcome: National cross-sectional survey in England and Wales}.
\newblock \emph{BMC Psychiatry}, 16.

\bibitem[{Winata et~al.(2017)Winata, Kampman, Yang, Dey, and Fung}]{winata2017nora}
Genta~Indra Winata, Onno~P. Kampman, Yang Yang, Anik Dey, and Pascale Fung. 2017.
\newblock Nora the empathetic psychologist.
\newblock In \emph{INTERSPEECH}, pages 3437--3438.

\bibitem[{Yang et~al.(2024)Yang, Tan, Sim, Lim, Tan, Kanneganti, Ooi, and Ong}]{yang_stress_2024}
Suyi Yang, Germaine Ke~Jia Tan, Kang Sim, Lucas Jun~Hao Lim, Benjamin Yong~Qiang Tan, Abhiram Kanneganti, Shirley Beng~Suat Ooi, and Lue~Ping Ong. 2024.
\newblock \href {https://doi.org/10.1371/journal.pone.0296798} {Stress and burnout amongst mental health professionals in {Singapore} during {Covid}-19 endemicity}.
\newblock \emph{PLOS ONE}, 19(1).

\bibitem[{Yeo et~al.(2024)Yeo, Qin, Ang, Chia, Ho, Ho, and Car}]{Yeo2024PrevalenceAC}
Pearlie Mei~En Yeo, Vicky~Mengqi Qin, Chin-Siang Ang, Michael Chia, Ringo Moon-Ho Ho, Andy Hau~Yan Ho, and Josip Car. 2024.
\newblock \href {https://api.semanticscholar.org/CorpusID:267657046} {Prevalence and correlates of depressive symptoms among matriculated university students in singapore during covid-19 pandemic: Findings from a repeated cross-sectional analysis}.
\newblock \emph{BMC Public Health}, 24.

\bibitem[{Yoon et~al.(2024)Yoon, Goh, Low, Weng, and Heaukulani}]{yoon2024user}
Sungwon Yoon, Hendra Goh, Xinyi~Casuarine Low, Janice~Huiqin Weng, and Creighton Heaukulani. 2024.
\newblock User perceptions and utilisation of features of an ai-enabled workplace digital mental wellness platform ‘mindline at work’.
\newblock \emph{BMJ Health \& Care Informatics}, 31(1).

\bibitem[{Young et~al.(2024)Young, Jawara, Nguyen, Daly, Huh-Yoo, and Razi}]{Young-etal-2024}
Jordyn Young, Laala~M Jawara, Diep~N Nguyen, Brian Daly, Jina Huh-Yoo, and Afsaneh Razi. 2024.
\newblock \href {https://doi.org/10.1145/3613904.3642574} {{The Role of AI in Peer Support for Young People: A Study of Preferences for Human- and AI-Generated Responses}}.
\newblock In \emph{Proceedings of the 2024 CHI Conference on Human Factors in Computing Systems}.

\bibitem[{Zhang et~al.(2022)Zhang, Schoene, Ji, and Ananiadou}]{zhang2022natural}
Tianlin Zhang, Annika~M Schoene, Shaoxiong Ji, and Sophia Ananiadou. 2022.
\newblock Natural language processing applied to mental illness detection: A narrative review.
\newblock \emph{NPJ digital medicine}, 5(1):1--13.

\bibitem[{Zhou et~al.(2018)Zhou, Gao, Li, and Shum}]{Zhou2018TheDA}
Li~Zhou, Jianfeng Gao, Di~Li, and Harry Shum. 2018.
\newblock \href {https://api.semanticscholar.org/CorpusID:56657857} {{The Design and Implementation of XiaoIce, an Empathetic Social Chatbot}}.
\newblock \emph{Computational Linguistics}, 46:53--93.

\end{thebibliography}

\appendix

\section{Example Conversations}
\label{sec:appendix}

\begin{figure*}[!t]
  \centering
  \includegraphics[width=\linewidth]{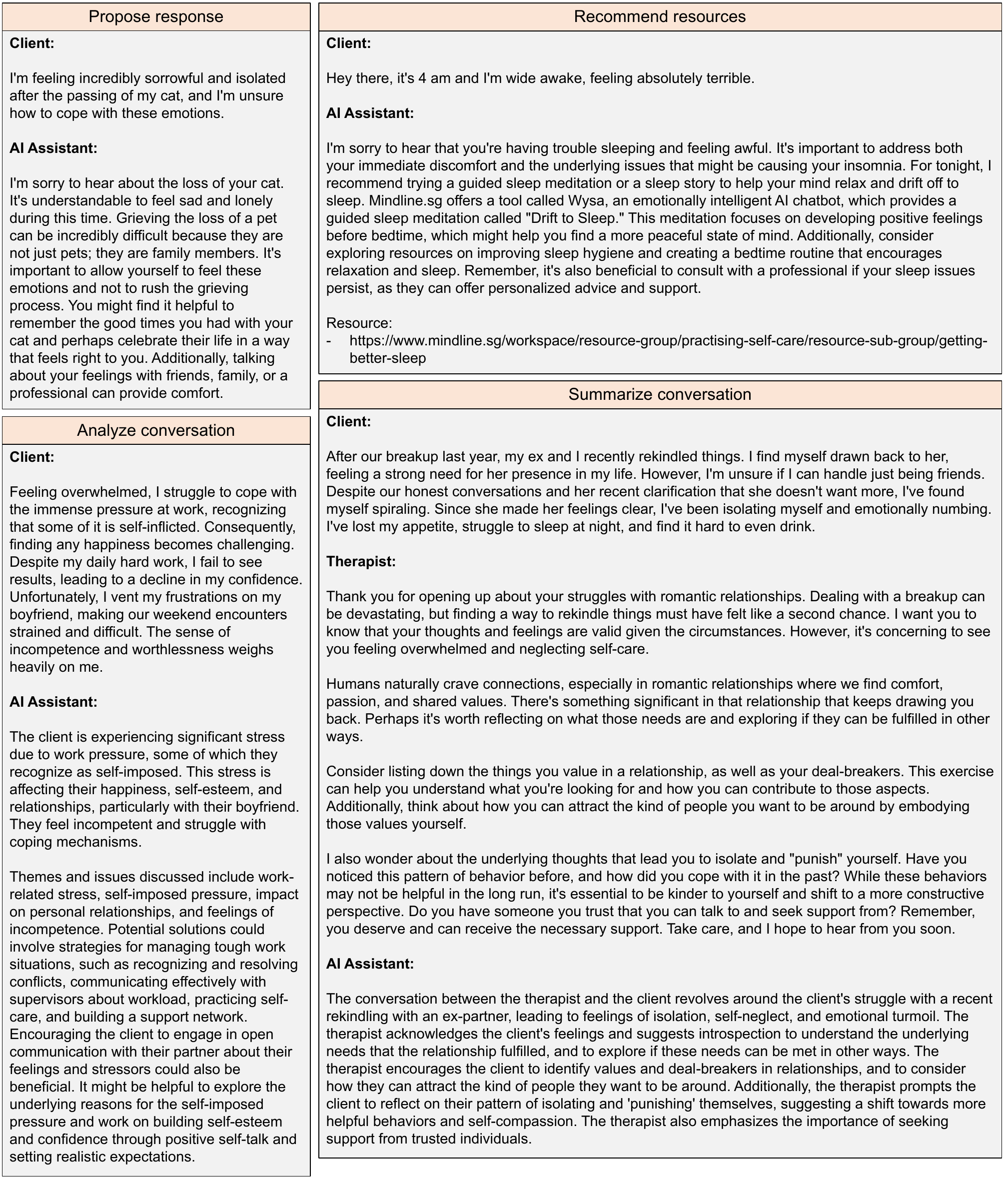}
  \caption{Examples of functionalities, including proposing responses, extracting relevant resources, and analyzing and summarizing conversations.}
  \label{fig:example-functionalities}
\end{figure*}

Figure~\ref{fig:example-functionalities} shows several examples of conversational functionalities.
We highlight several system capabilities.
For the \textbf{Propose Response} functionality, the AI assistant acknowledges the client's emotions and stresses the importance of a social support network.
For the \textbf{Recommend Resources} functionality, it briefly empathizes with the client, and then recommends a relevant sleep-related exercise, while also explaining why such exercises may be helpful.
For the \textbf{Analyze Conversation} functionality, the AI assistant extracts relevant themes from the client's message and discusses potential therapeutic directions, such as learning more about managing work-related stressors, encouraging open communication with their partner, and exploring self-esteem issues that may underlie the symptoms discussed.
While any trained therapist will understand these insights, this functionality can prove useful when meeting a new client with a chat history or when the therapist has not received extensive training.
For the \textbf{Summarize Conversation} functionality, the AI assistant condenses an exchange between client and therapist in a short paragraph by extracting the main themes and issues and discussing focal points of the healing process.

\begin{table*}[!ht]

    \centering
    \small
    \renewcommand{\arraystretch}{1.5}

    \begin{tabular}{p{2.5cm} | p{2cm} | p{10cm}}

        \toprule
        \textbf{Item} & \textbf{Label} & \textbf{Description} \\
        \midrule
        Concern &
        \verb|concern| &
        A therapist conveys concern by showing a regard for and interest in the client. The therapist seems engaged and involved with the client and attentive to what the client has said. The therapist’s voice has a soft resonance that supports and enhances the client’s concerned expressions. \\
        Expressiveness &
        \textit{N/A (unused in a text-based environment)} &
        A therapist’s voice demonstrates expressiveness when the therapist speaks with energy and varies the pitch of his or her voice to accommodate the mood or disposition of the client.
        \\
        Resonate or capture client feelings &
        \verb|resonate| &
        A therapist resonates with or captures the intensity of the client’s feelings when he or she speaks with a tone and emphasis that matches the client’s emotional state or pitches words or phrases that underscore how the client feels.
        \\
        Warmth &
        \verb|warmth| &
        A therapist resonates with or captures the intensity of the client’s feelings when he or she speaks with a tone and emphasis that matches the client’s emotional state or pitches words or phrases that underscore how the client feels.
        \\
        Attuned to the client’s inner world &
        \verb|attuned| &
        A client’s inner world is defined as the client’s feelings, perceptions, memories, meanings, bodily sensations, and core values. A therapist is attuned to a client’s inner world when they provide moment-to-moment verbal acknowledgment of the client’s expressions. These acknowledgments suit, agree with, or support the client's mood and reflections. The therapist is attentive to nuances of meaning and feeling conveyed in a client’s statements beyond surface content and shows a genuine understanding of the client’s inner world.
        \\
        Understanding cognitive framework &
        \verb|cognitive| &
        A therapist demonstrates an understanding of the client’s cognitive framework and meanings when he or she clearly follows what the client has said and accurately reflects this understanding to the client. In short, the therapist and client are on the same page. The therapist is careful to provide ample opportunities for the client to state his or her views in order to permit the fullest and most accurate understanding of the client. The interaction conveys that the therapist values knowing what the client means or intends by his or her statements without predetermination or judgment.
        \\
        Understanding feelings/inner experience &
        \verb|understanding| &
        A therapist conveys an understanding of a client’s feelings and inner experience when they show a sensitive appreciation and gentle caring for the client’s emotional state. A therapist provides ample opportunities for the client to explore his or her emotional reactions. The therapist accurately reflects how the client feels by appropriately labeling feeling states with words (e.g., anger, sadness, frustration, etc.) or metaphors (e.g., “It’s as if you are pent up and feel about to explode”) to clarify and crystallize for the client what they are experiencing emotionally.
        \\
        Acceptance of feelings/inner experiences &
        \verb|acceptance| &
        A therapist shows acceptance of the client’s feelings and inner experience when he or she validates the client’s experience and reflects the client’s feelings without judgment or a dismissive attitude. The therapist is unconditionally open to and respectful of how the client feels. The therapist’s stance is one of genuineness and honesty instead of seemingly feigning concern or appearing inauthentic.
        \\
        Responsiveness &
        \textit{N/A (unused in a text-based environment)} &
        A therapist shows responsiveness to the client by adjusting his or her responses to the client’s statements or nonverbal communications during the conversation. The therapist follows the client’s lead in the conversation instead of trying to steer the discussion to the therapist’s agenda or interests.
    \end{tabular}
    
    \caption{
        Adapted TES rating items and their descriptions~\citep{Decker2013TES}.
    }
    \label{tab:tes-rubrics}
\end{table*}

\begin{figure}[!ht]

    \includegraphics[width=\linewidth]{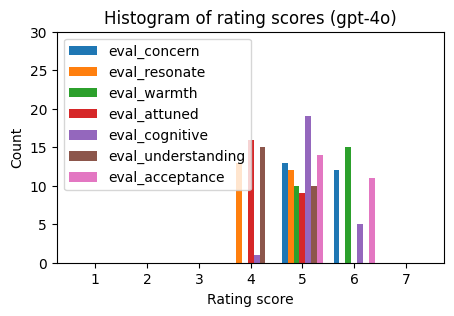}
    \includegraphics[width=\linewidth]{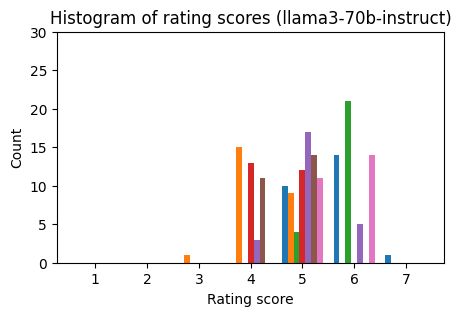}
    \includegraphics[width=\linewidth]{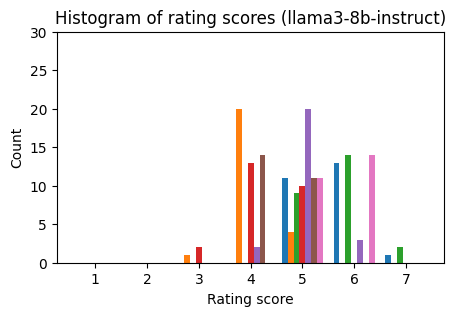}
    \includegraphics[width=\linewidth]{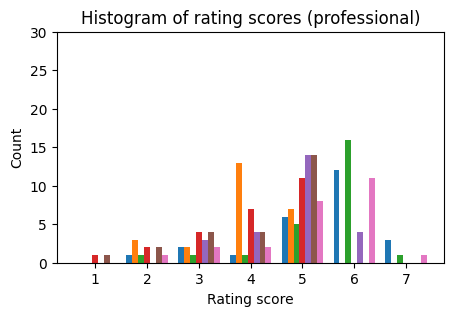}

    \caption{Histograms of rating items for responses generated by GPT-4o (first row), Llama 3 70b (second), Llama 3 8b (third), and professional therapists (last), as evaluated by GPT-4o.}
    \label{fig:eval-histogramsGPT}
\end{figure}

\begin{figure}[!ht]

    \includegraphics[width=\linewidth]{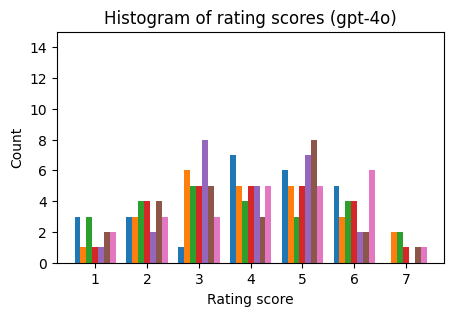}
    \includegraphics[width=\linewidth]{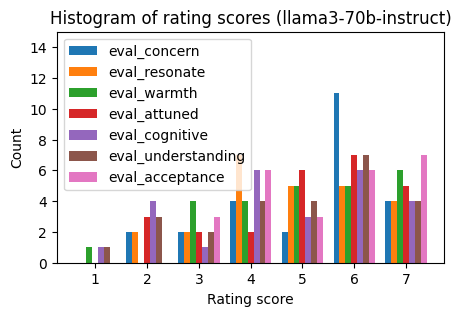}
    \includegraphics[width=\linewidth]{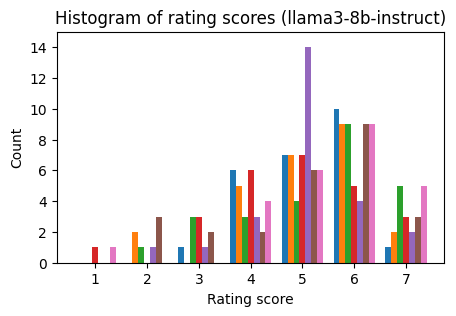}
    \includegraphics[width=\linewidth]{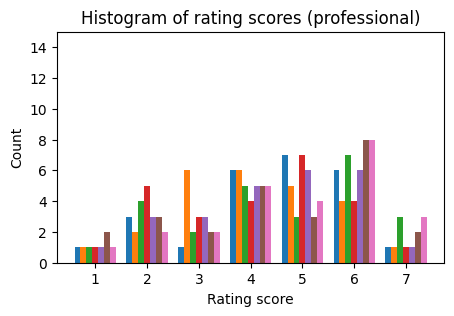}

    \caption{Histograms of rating items for responses generated by GPT-4o (first row), Llama 3 70b (second), Llama 3 8b (third), and professional therapists (last), as evaluated by human raters.}
    \label{fig:eval-histogramsHuman}
\end{figure}

\end{document}